\documentclass[aps,pre,twocolumn,showpacs,superscriptaddress,groupedaddress]{revtex4-1}  % for review and submission
\usepackage{graphicx}  % needed for figures
\usepackage{dcolumn}   % needed for some tables
\usepackage{bm}        % for math
\usepackage{amssymb}   % for math

\begin{document}

% The following information is for internal review, please remove them for submission

% the following line is for submission, including submission to the arXiv!!
%\hspace{5.2in} \mbox{Fermilab-Pub-04/xxx-E}

\title{Akhmediev breathers, Ma solitons and general breathers from rogue waves: A case study in Manakov system}
\author{N. Vishnu Priya, M. Senthilvelan and M. Lakshmanan}
\address{Centre for Nonlinear Dynamics, School of Physics, Bharathidasan University, Tiruchirappalli - 620 024, Tamil Nadu, India.}
%\input author_list.tex       % D0 authors (remove the first 3 lines
                             % of this file prior to submission, they
                             % contain a time stamp for the authorlist)
                             % (includes institutions and visitors

\begin{abstract}
We present explicit forms of general breather (GB), Akhmediev breather (AB), Ma soliton (MS) and rogue wave (RW) solutions of the two component nonlinear Schr\"{o}dinger (NLS) equation, namely Manakov equation.  We derive these solutions through two different routes.  In the forward route we first construct a suitable periodic envelope soliton solution to this model from which we derive GB, AB, MS and RW solutions.  We then consider the RW solution as the starting point and derive AB, MS and GB in the reverse direction.  The second approach has not been illustrated so far for the two component NLS equation.  Our results show that the above rational solutions of the Manakov system can be derived from the standard scalar nonlinear Schr\"{o}dinger equation with a modified nonlinearity parameter.  Through this two way approach we establish a broader understanding of these rational solutions which will be of interest in a variety of situations. 
\end{abstract}

\maketitle

%\section{\label{sec:level1}First-level heading}
% sections are not used for PRL papers

\section{Introduction} 
We consider the integrable system of two coupled nonlinear Schr\"{o}dinger equations (NLSEs), namely the Manakov system,
\begin{eqnarray}
\label{cnls01}
\nonumber iq_{1t}+q_{1xx}+2\mu(|q_1|^2+|q_2|^2)q_1&=&0,\\
iq_{2t}+q_{2xx}+2\mu(|q_1|^2+|q_2|^2)q_2&=&0,
\end{eqnarray}
where $q_1$ and $q_2$ are wave envelopes, $x$ and $t$ are space and time variables respectively, $\mu$ is a real constant and subscripts denote partial derivatives with respect to the corresponding variables.
Eq.(\ref{cnls01}) represents the propagation of an optical pulse in a birefringent optical fibre and in wavelength division multiplexed system \cite{Manakov}.  It has been studied widely in the literature \cite{{Manakov},{Menyuk},{Kodama},{Wright},{Park},{Radha},{Radha1}}.  The complete integrability of this system of coupled NLSEs was first established by Manakov \cite{Manakov}. 
It has also been demonstrated that this two-component vector generalization of the focusing NLS equation admits several interesting properties including (i) infinite number of conservation laws \cite{Manakov},  (ii) Lax pair \cite{Manakov}, (iii) an infinite-dimensional algebra of non-commutative symmetries \cite{Kodama}, (iv) bilinear representation and bright multi-soliton solutions \cite{Radha} and so on.  Eq.(\ref{cnls01}) also appears in multi-component Bose-Einstein condensates \cite{Busch}, bio-physics \cite{Scott}, finance \cite{Finance1} and oceanographic studies \cite{Ocean}.  Solitons in coupled NLSEs have been the subject of intense study over the past few years because of their interesting collision properties and their robustness against external perturbations.  The explicit multi-bright and dark soliton solutions of (\ref{cnls01}) were obtained by Radhakrishnan et al.\cite{Radha,Radha1}.  It has also been demonstrated that the soliton solutions of (\ref{cnls01}) exhibit a fascinating shape changing collision, resulting in a redistribution of intensity between the modes of the two solitons, which is not observed in the scalar NLSE case \cite{Radha1}.

\par Very recently, a new rational solution called rogue wave (RW) solution has attracted considerable attention \cite{{Pelin},{Akhmediev},{BEC},{Capillary},{Plasma},{Finance},{Akhmediev1},{Akhmediev2},{Kalla},{Zhai},{Ling},{Degasperis},{Solli},{Efimov},{Fedele},{Smirnov},{Shukla},{Ohta},{Dysthe},{Akhmediev3},{Akhmediev4},{Akhmediev5},{Akhmediev6},{Akhmediev7},{Akhmediev8},{Akhmediev9}}.  RWs, alternatively called freak or giant waves, were first observed in arbitrary depth of ocean circumstances.  A wave is classified into this category when its wave height (distance from trough to crest) reaches a value which is atleast twice that of the significant wave height \cite{Pelin}.  These waves may arise from the instability of a certain class of initial conditions that tend to grow exponentially and thus have the possibility of increasing up to very high amplitudes, due to modulation instability \cite{Akhmediev}.  Over the years RWs have also been observed in models that arise in the description of multi-component Bose-Einstein condensates \cite{BEC}, capillary waves \cite{Capillary}, multi-component plasmas \cite{Plasma} and even in finance \cite{Finance}.  Recently efforts have been made to explain the RW excitation through a nonlinear process.  It has been found that the NLS equation can describe many dynamical features of the RW.  Certain kinds of exact solutions of NLS equation have been considered to describe possible mechanism for the formation of RWs such as Peregrine soliton, time periodic breather or Ma soliton (MS) and space periodic breather or Akhmediev breather (AB) \cite{{Akhmediev1},{Akhmediev2}}.
As a consequence attempts have been made to construct RW solution through different methods for the NLS equation and its higher derivative generalizations.  One way of obtaining RW solution or Peregrine soliton for a given system is to first construct a breather solution, either AB or MS.  From the latter, the RW solution can be deduced in an appropriate limit. 
 
\par As far as the system of two coupled NLSEs is concerned, in recent years, the following studies have been undertaken.  Breathers and rational breather solutions of multi-component NLSE are presented in \cite{Kalla} in a determinantal form as limiting cases in suitable degenerations of algebro-geometric solutions. Explicit first, second and third order RW solutions of (\ref{cnls01}) have been constructed in Ref.\cite*{Zhai} through the modified Darboux transformation method. The authors have also studied some basic properties of multi-rogue wave  solutions and their collision structures.  In Ref.\cite*{Ling} two types of RW solutions through Darboux transformation method have been derived.  The authors have shown that while the first kind of RW solution is similar to the first-order RW solution of NLSE, the second kind of RW behaves differently from that of the first order rogue wave solution.  Recently, an in-depth analysis on the construction of vector Peregrine soliton solution and bright-dark-RW solution of Eq.(\ref{cnls01}) has been made in Ref.\cite*{Degasperis}.  In all the above cited works only the explicit forms of RWs are given and we do not see any simple tractable form of breather solutions.     

\par Since the breather solution plays an important role in the formation of RWs we aim here to derive the breather solution for the widely studied nonlinear evolution equation (\ref{cnls01}).  We divide our analysis into two parts.  In the first part, we construct a periodic envelope two soliton solution through Hirota's bilinearization method.  By appropriately restricting the wave number (of one of the solitons to be the complex conjugate of the other) which appears in the two soliton solution we obtain the GB form of (\ref{cnls01}).  From the GB solution we derive AB, MS and RW solutions.  We note here that through a restricted set of transformations, the GB solution of the Manakov equation can be obtained from the GB solution of the NLS equation with a modified nonlinear parameter.  In other words one can generate the above rational solutions of the Manakov system from the standard nonlinear Schr\"{o}dinger equation with a modified nonlinear parameter.

\par In the second part of our work we analyze the reverse problem: How can one construct a AB or MS or GB from a RW solution?  We answer this question by rewriting the RW solution in a factorized form and then generalizing this factorized form in an imbricate series expression \cite{{Book},{Boyd},{Toda}} with certain unknown parameters in this series, following the earlier work of Tajiri and Watanabe \cite{Tajiri} for the case of scalar NLS equation and finally finding these unknown parameters in the imbricate series by substituting it in Eq. (\ref{cnls01}) and solving the resultant equations.  With three different forms of the imbricate series we derive the AB, MS and GB solutions from the RW solution of (\ref{cnls01}). 

\par The plan of the paper is follows.  In the following Sec. \ref{GB} we construct the explicit form of the GB solution of the two coupled NLSE system (\ref{cnls01}) through Hirota's bilinearization method.  We then explain the method of deriving AB, MS and RW solutions from the GB solution.  The obtained form of RW solution coincides with the ones in the literature.  In Sec. \ref{RWtoAB}, we discuss the method of constructing the AB solution from the RW solution.  In Sec. \ref{RWtoMS}, we demonstrate the construction of MS from RW.  In Sec. \ref{RWtoGB}, we formulate the imbricate series form for the RWs with certain unknown arbitrary functions in it and then compare this expression with the one derived from the GB in the same way.  The comparison provides exact expressions for the unknown arbitrary functions which appear in the imbricate series of the RW.  In this way we establish a method of constructing GBs from RW.  Finally, in Sec. \ref{conclusions} we present our conclusions.  

\section{General Breathers} 
\label{GB}
We seek a periodic envelope solution to the CNLS equations (\ref{cnls01}) with the boundary conditions $|q_i|^2 \to \tau_i^2$, i=1,2, as $x \to \pm\infty$, where $\tau_1$ and $\tau_2$ are real constants.  To start with we bilinearize Eq.(\ref{cnls01}) through the transformation $q_1=\frac{g}{f}$ and $q_2=\frac{h}{f}$, where $g$ and $h$ are complex functions and $f$ is a real function.  The resultant bilinearized forms read
\begin{eqnarray}
(iD_t+2ikD_x+D_x^2)g.f=0,\nonumber\\
(iD_t+2ikD_x+D_x^2)h.f=0,\nonumber\\
(D_t^2+2\mu (\tau_1^2+\tau_2^2))f.f-2\mu(|g|^2+|h|^2)=0.
\label{pct2}
\end{eqnarray}
\par In the above, $D_t$ and $D_x$ are Hirota's bilinear operators \cite{Radha}.  Once the nonlinear evolutionary equation has been bilinearized, with truncated parameter expansion at different levels, a series of solutions, in particular the N-soliton solution, can be obtained.  As far as Eqs.(\ref{pct2}) are concerned, the N-soliton solution, can be obtained with respect to the expansion parameter $\chi$, that is $g=g_0(1+\chi g_1+\chi^2 g_2+...)$, $h=h_0(1+\chi h_1+\chi^2 h_2+...)$ and $f=(1+\chi f_1+\chi^2 f_2+...)$, where $g_i$'s and $h_i$'s, $i=0,1,2,..N$, are complex functions of $x$ and $t$ and $f_i$'s are real variables.  
\par As our aim is to obtain the two soliton solution we terminate the expansion at quadratic powers in $\chi$, that is $g=g_0(1+\chi g_1+\chi^2g_2)$, $h=h_0(1+\chi h_1+\chi^2h_2)$ and $f=(1+\chi f_1+\chi^2f_2)$.  The resultant two soliton solution emerges in the form
\begin{eqnarray}
q_1=\tau_1e^{i\theta}\frac{g}{f} \ \text{and} \ q_2=\tau_2e^{i\theta}\frac{h}{f}, \ \theta=kx-\omega t,
\label{pct3}
\end{eqnarray}
where
\begin{eqnarray}
g&=&h=1+e^{\eta_1+2i\phi_1}+e^{\eta_2+2i\phi_2}+a e^{\eta_1+\eta_2+2i\phi_1+2i\phi_2},\nonumber\\
f&=&1+e^{\eta_1}+e^{\eta_2}+e^{\eta_1+\eta_2}, \ \eta_j=p_jx-\Omega_jt+\eta_j^0, \ j=1,2.\nonumber
\end{eqnarray}
In the above $p_j$, $\Omega_j$, $\eta_j^0$ and $\phi_j$, $j=1,2$, are complex parameters and
\begin{eqnarray}
\omega&=&k^2-2\mu(\tau_1^2+\tau_2^2), \ p_j=2i\sqrt{\mu(\tau_1^2+\tau_2^2)}\sin\phi_j,\nonumber\\
\Omega_j&=&2k_jp_j-p_j^2\cot\phi_j, \ j=1,2, \nonumber\\
a&=&\left(\frac{\sin\frac{1}{2}(\phi_1-\phi_2)}{\sin\frac{1}{2}(\phi_1+\phi_2)}\right)^2.
\label{pct4}
\end{eqnarray}
In the above we have chosen $g=h$.  One can proceed by assuming $g\neq h$ also.  However, in order to obtain the required breather solutions we find that one has to essentially fix $g=h$.  
\par We have not pursued the possibility of a more general bilinearization than Eq. (\ref{pct2}) which will lead to the possibility $g\neq h$ in (\ref{pct3}).  We hope to consider such a generalization in future.  So in our analysis we have made this choice in the beginning itself.
\par Note that due to the choice $g=h$ in Eq. (\ref{pct3}) one can effectively make a transformation
\begin{eqnarray}
q_1=\tau_1q, \; \; q_2=\tau_2q,
\label{pc4}
\end{eqnarray}
so that 
\begin{eqnarray}
q=e^{i\theta}\frac{g}{f}=e^{i\theta}\frac{h}{f}
\label{pc5}
\end{eqnarray}
satisfies the scalar NLS equation with the nonlinearity parameter $2\mu(\tau_1^2+\tau_2^2)$:
\begin{eqnarray}
iq_t+q_{xx}+2\mu(\tau_1^2+\tau_2^2)|q|^2q=0.
\label{pc6}
\end{eqnarray}
Consequently one can write down the breather solution of the variables $q_1$ and $q_2$ equivalently from the breather solution of the above scalar NLS equation as well.
\par The constants, $\phi_j=\phi_{jR}+i\phi_{jI}\neq 0$, $j=1,2,$ help us to split the above breather expression into Akhmediev and Ma breathers as we see below.  To obtain the breather solution from the above two soliton solution we take $\eta_1=\eta_2^*\equiv \eta$ and $\phi_2=\phi_1^*\pm \pi$.  Substituting these two restrictions in (\ref{pct3}) and considering $\eta=\eta_{R}+i\eta_{I}$ and  $\phi_1=\phi_{R}+i\phi_{I}$, the exponential functions appearing in (\ref{pct3}) can be rewritten in terms of trigonometric and hyperbolic functions.  The resultant expression for $q$ turns out to be                                                                                                          
\begin{eqnarray}
q=&&\cos2\phi_R e^{i(\theta+2\phi_R)}\bigg[1+\frac{1}{\sqrt{a}\cosh(\eta_R+\sigma)+\cos\eta_I}\nonumber\\
&&\times\bigg(\bigg(\frac{\cos2\phi_I}{\cos2\phi_R}-1\bigg)\cos\eta_I+i\bigg(\sqrt{a}\tan2\phi_R\nonumber\\
&&\times\sinh(\eta_R+\sigma)-\frac{\sinh2\phi_I}{\cos2\phi_R}\sin\eta_I\bigg)\bigg)\bigg],
\label{pct5}
\end{eqnarray}
where $\eta_R=p_Rx-\Omega_Rt+\eta_R^0$,  $\eta_I=p_Ix-\Omega_It+\eta_I^0$, $p_1=p_R+ip_I$, $\Omega_1=\Omega_R+i\Omega_I$, $\eta_R^0$, $\eta_I^0$ and $\sigma$ are constants.  The exact forms of $p_R$, $\Omega_R$, $p_I$ and $\Omega_I$ are given below:
\begin{eqnarray}
p_R&=&-2\sqrt{\mu(\tau_1^2+\tau_2^2)}\cos\phi_R\sinh\phi_I, \nonumber\\ p_I&=&2\sqrt{\mu(\tau_1^2+\tau_2^2)}\sin\phi_R\cosh\phi_I,\nonumber\\
\Omega_R&=&2kp_R-\frac{(p_R^2-p_I^2)\sin 2\phi_R+2p_Rp_I\sinh 2\phi_I}{\cosh 2\phi_I-\cos 2\phi_R}, \nonumber\\ \Omega_I&=&2kp_I-\frac{(p_R^2-p_I^2)\sinh 2\phi_I+2p_Rp_I\sin 2\phi_R}{\cosh 2\phi_I-\cos 2\phi_R}.\nonumber
\end{eqnarray}
Consequently the solution of the Manakov equation (\ref{cnls01}) can be obtained from (\ref{pc4}) as $q_1=\tau_1q$ and $q_2=\tau_2q$.
Expression (\ref{pct5}) combined with (\ref{pc4}) constitutes the GB solution of the CNLS equations (\ref{cnls01}).  Fig. 1 illustrates the behavior of this breather solution, which is periodic both in space and in time.  From the GB solution we can derive AB, MS and RW solutions by restricting the parameters $\phi_R$ and $\phi_I$ suitably.  In the following, we report the explicit forms of these solutions.  
\subsection{AB from GB}
To derive the AB solution, we consider the choice $\phi_R\neq 0$ and $\phi_I=0$ in (\ref{pct5}).  This restriction fixes the wave number to be pure imaginary.  In this case we find
\begin{eqnarray}
p_R&=&0, \ p_I=2\sqrt{\mu(\tau_1^2+\tau_2^2)}\sin\phi_R,\nonumber\\
\Omega_R&=&p_I^2\cot\phi_R, \ \Omega_I=2kp_I.
\label{pct6}
\end{eqnarray}
Substituting the GB solution (\ref{pct5}) in (\ref{pc4}), the latter provides
\begin{eqnarray}
q_1&=&\tau_1\cos(2\phi_R)e^{i(\theta+2\phi_R)}(1+L), \nonumber\\ q_2&=&\tau_2\cos(2\phi_R)e^{i(\theta+2\phi_R)}(1+L),\label{pct7}\\
L&=&\frac{(\frac{1}{\cos(2\phi_R)}-1)\cos\eta_I+i\sqrt{a}\tan(2\phi_R)\sinh(\eta_R+\sigma)}{\sqrt{a}\cosh(\eta_R+\sigma)+\cos\eta_I}.\nonumber\\
\nonumber
\end{eqnarray} 
Here $\eta_R=-\Omega_Rt+\eta_R^0$ and $\eta_I=p_Ix-\Omega_It+\eta_I^0$.
We have plotted the solution (\ref{pct7}) in Fig. 2.  The solution is periodic in $x$ and localized in $t$.  This spatially periodic breather is nothing but the AB solution.
   
\subsection{MS from GB}
Now we consider the other case, $\phi_R=0$ and $\phi_I\neq 0$ with $k=0$ in the GB solution ((\ref{pc4}) with (\ref{pct5})).  This restriction fixes the imaginary part of the wave number to be zero.  This in turn provides another form of the breather solution which will propagate only in the time direction, that is
\begin{eqnarray}
q_1&=&\tau_1e^{i\theta}(1+V), \ q_2=\tau_2e^{i\theta}(1+V),\label{pct8}\\
V&=&\frac{\cosh(2\phi_I)-1}{\sqrt{a}\cosh(\eta_R+\sigma)+\cos\eta_I}(\cos\eta_I-i\sinh(2\phi_I)\sin\eta_I),\nonumber
\end{eqnarray}
where
\begin{eqnarray}
p_R&=&-2\sqrt{\mu(\tau_1^2+\tau_2^2)}\sinh\phi_I, \ p_I=0, \nonumber\\ \Omega_R&=&0,
\Omega_I=p_R^2\coth\phi_I, \ \eta_R=p_Rx-\Omega_Rt+\eta_R^0 \nonumber
\end{eqnarray}
\vspace{-.3cm}
and
\vspace{-.3cm}
\begin{eqnarray}
\hspace{-5.2cm}\eta_I=-\Omega_It+\eta_I^0.
\label{pct9}
\end{eqnarray}
We depict this solution in Fig. 3.  The plot confirms that the solution is periodic in $t$ and localized in $x$.  The wave solution which is temporally breathing and spatially oscillating is called a Ma breather/MS.   

\subsection{RW from GB}
To construct the RW solution from the GB expression we consider both $\phi_R$ and $\phi_I$ to be non-zero and evaluate the latter in the limit $\epsilon\to 0$ by implementing a Taylor expansion.  Doing so we find $\phi_R=\epsilon\gamma$ and $\phi_I=\epsilon\rho$, where $\epsilon$ is a small parameter, $\gamma$ and $\rho$ are constants.  Substituting these two expressions in (\ref{pct4}) with the restriction $\eta_2=\eta_1^*$, $\phi_2=\phi_1^*+\pi$ and making the Taylor expansion at $\epsilon\to 0$, we obtain
\begin{eqnarray}
p_R&=&-2\sqrt{\mu (\tau_1^2+\tau_2^2)}\rho\epsilon+O(\epsilon^3),\nonumber\\
p_I&=&2\sqrt{\mu (\tau_1^2+\tau_2^2)}\gamma\epsilon+O(\epsilon^3),\nonumber\\
\Omega_R&=&(4\mu(\tau_1^2+\tau_2^2)\gamma-4k\sqrt{\mu (\tau_1^2+\tau_2^2)}\rho)\epsilon+O(\epsilon^3),\nonumber\\
\Omega_I&=&(4\mu(\tau_1^2+\tau_2^2)\rho+4k\sqrt{\mu (\tau_1^2+\tau_2^2)}\gamma)\epsilon+O(\epsilon^3),\nonumber\\
\sqrt{a}&=&1+\frac{1}{2}(\gamma^2+\rho^2)\epsilon^2,\label{pct10}\\
f&=&((\tilde{\eta}_R^2+\tilde{\eta}_I^2)+(\gamma^2+\rho^2))\epsilon^2+O(\epsilon^3),\nonumber\\
g&=&((\tilde{\eta}_R^2+\tilde{\eta}_I^2)-3(\gamma^2+\rho^2)+4i(\gamma \tilde{\eta}_R+\rho\tilde{\eta}_I))\epsilon^2+O(\epsilon^3).\nonumber\\
\nonumber
\end{eqnarray}
It is also straightforward to check that $\eta_R-\eta_R^0=\epsilon\tilde{\eta}_R+O(\epsilon^2)$ and $\eta_I-\eta_I^0=\epsilon\tilde{\eta_I}+O(\epsilon^2).$  Substituting the above expressions, (\ref{pct10}), in the general breather form (\ref{pct5}) and taking the limit $\epsilon\to 0$ in the resultant expression, we arrive at
\begin{eqnarray}
q_1&=&\tau_1e^{i\theta}(1-Q) \ \text{and} \ q_2=\tau_2e^{i\theta}(1-Q),\label{pct11}
\end{eqnarray}
where
\begin{eqnarray}
Q&=&\frac{4+16i\mu(\tau_1^2+\tau_2^2)t}{1+4\mu(\tau_1^2+\tau_2^2)(x-2kt)^2+16\mu^2(\tau_1^2+\tau_2^2)^2t^2},\nonumber\\
\nonumber
\end{eqnarray} 
which is nothing but the RW solution of CNLS system which is localized both in space and time.  The RW solution of the CNLS equation matches with the one presented in \cite{Zhai}.  We note that the restriction $\tau_2=0$ in (\ref{pct8}) provides the RW solution of the scalar NLS equation.  A typical evolution of the RW is shown in Fig. 4.
\begin{figure}
\includegraphics[width=\linewidth]{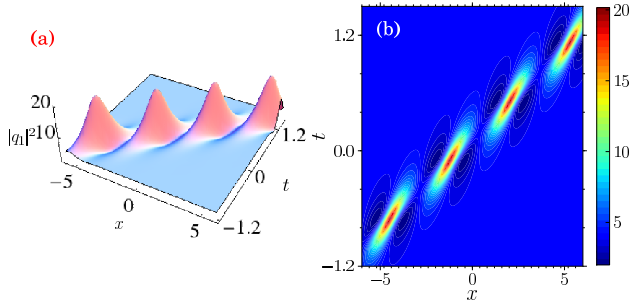}
\caption{(Color online) (a) General breather profile of $q_1$ for the values $\tau_1 = 2$, $\tau_2=1$, $\phi_R = 4$, $\phi_I = 1$, $\eta_I^o=0.5$, $\eta_R^o=0.2$, $k=0.24$, $\mu=0.2$ in Eqs. (\ref{pc4}) and (\ref{pct5}).  (b) Corresponding contour plot.  Similar profile occurs for $q_2$ also (not shown here).}
\end{figure}
\begin{figure}
\includegraphics[width=1\linewidth]{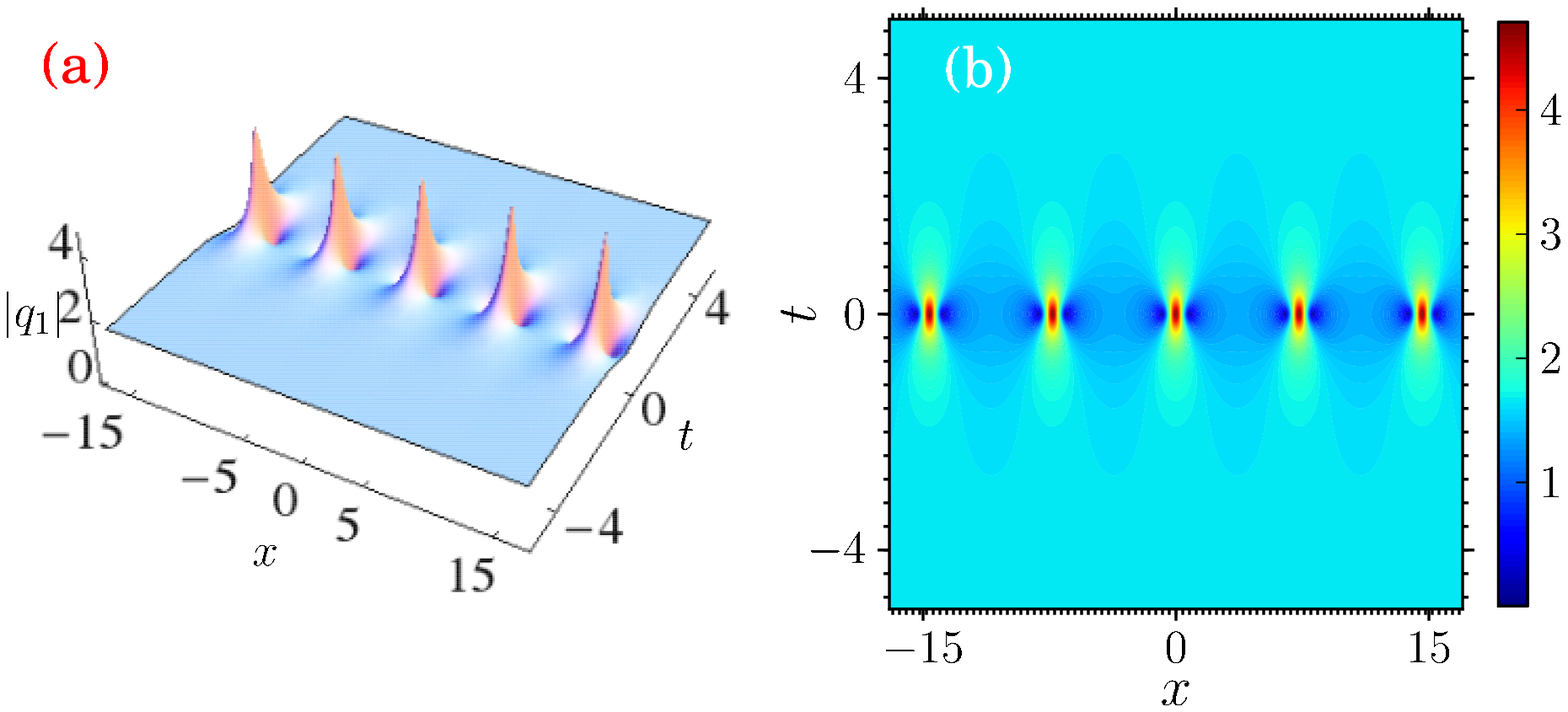}
\caption{(Color online) (a) Akhmediev breather profile of $q_1$ for the values $\tau_1=2$, $\tau_2=1$, $\phi_R=0.5$, $\mu=0.2$, $\eta_I^o=0.5$, $\eta_R^o=0.1$ in Eq. (\ref{pct7}).  (b) Corresponding contour plot.  Similar profile occurs for $q_2$ also (not shown here).}
\end{figure}
\begin{figure}
\includegraphics[width=1\linewidth]{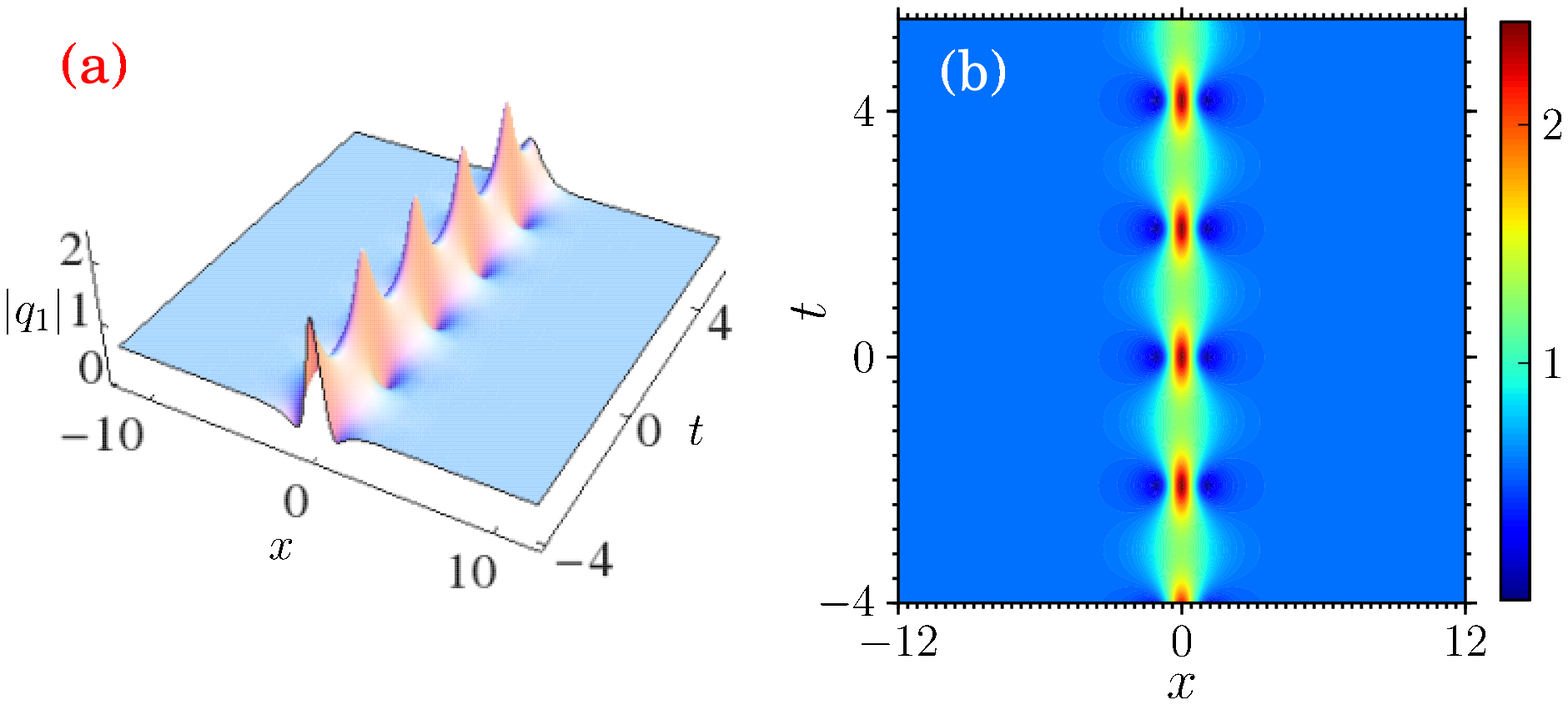}
\caption{(Color online) (a) Ma breather profile of $q_1$ for the values $\tau_1=2$, $\tau_2=1$, $\phi_I=0.8$, $\mu=0.2$, $\eta_I^o=0.4$, $\eta_R^o=0.3$ in Eq. (\ref{pct8}).  (b) Corresponding contour plot.  Similar profile occurs for $q_2$ also (not shown here).}
\end{figure}
\begin{figure}
\includegraphics[width=\linewidth]{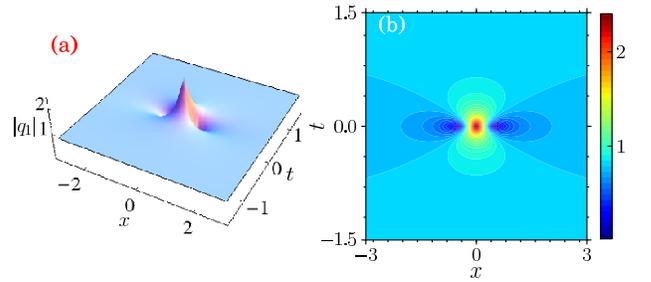}
\caption{(Color online) (a) Rogue wave profile of $q_1$ for the values $\tau_1=0.8$, $\tau_2=1.5$, $\mu=1$, $k=0$ in Eq. (\ref{pct11}).  (b) Corresponding contour plot.  Similar profile occurs for $q_2$ also (not shown here).}
\end{figure}
\par In the above, we derived AB, MB and RW solutions from the GB solution.  On the other hand we now point out the interesting possibility that one can also construct the above solutions from the RW solution itself in a reverse way.  In the following, we demonstrate this by following the procedure of Tajiri and Watanabe for the case of the scalar NLS equation \cite{Tajiri}.  To do so we consider the RW solutions in an imbricate series form. 
%%%%%%%%%%%%%%%%%%%%%%%%%%%%%%%%%%%%%%%%%%%%%%%%%%%%%%%%%%%%%%%%%%%%%%%%%%%%%%%%%%%%%%%%%%%%%%%%%%%%%%%%%%%%%%%%%%%%%%%%%%%%%%%%5%%
\section{AB from RW}
\label{RWtoAB}
To derive AB from RW solution we first factorize the RW solution (\ref{pct11}) in the following form, namely
\begin{eqnarray}
q_1=&&\hspace{.2cm}\tau_1\exp(i(kx-(k^2-2\mu(\tau_1^2+\tau_2^2))t))\nonumber\\
&&\hspace{-.7cm}\times\bigg(1+\frac{1}{2i\mu(\tau_1^2+\tau_2^2)t+\frac{1}{2}\sqrt{1+4\mu(\tau_1^2+\tau_2^2)(x-2kt)^2}}\bigg)\nonumber\\
&&\hspace{-.7cm}\times\bigg(1+\frac{1}{2i\mu(\tau_1^2+\tau_2^2)t-\frac{1}{2}\sqrt{1+4\mu(\tau_1^2+\tau_2^2)(x-2kt)^2}}\bigg),\nonumber
\end{eqnarray}  
\begin{eqnarray}
q_2=&&\hspace{.2cm}\tau_2\exp(i(kx-(k^2-2\mu(\tau_1^2+\tau_2^2))t))\nonumber\\
&&\hspace{-.7cm}\times\bigg(1+\frac{1}{2i\mu(\tau_1^2+\tau_2^2)t+\frac{1}{2}\sqrt{1+4\mu(\tau_1^2+\tau_2^2)(x-2kt)^2}}\bigg)\nonumber\\
&&\hspace{-.7cm}\times\bigg(1+\frac{1}{2i\mu(\tau_1^2+\tau_2^2)t-\frac{1}{2}\sqrt{1+4\mu(\tau_1^2+\tau_2^2)(x-2kt)^2}}\bigg).\nonumber\\
\label{pct12}
\end{eqnarray}
We note here that one of the remarkable properties displayed by many classical nonlinear evolution equations possessing soliton modes is a nonlinear superposition principle \cite{{Book},{Boyd},{Toda}}.  More precisely, an infinite array of solitons placed at equal intervals constitutes an exact periodic solution of the evolution equations.  For example, if we take the algebraic soliton solution of the modified Korteweg-de Vries equation in the form $u=u_o-\frac{4u_0}{4u_0^2(x-6u_0^2t)^2+1}$, we can write a more general solution through a superposition of these algebraic solitons as $u=u_0-\sum_{m=-\infty}^{\infty}\frac{4u_0}{4u_0^2(x-m\lambda-ct)^2+1}$, where $\lambda$ is the spacing between successive peaks of the sequence of solitary pulses and $c$ is the phase speed of the pattern, which is to be determined.  By rewriting the algebraic soliton solution as a hyperbolic cot function and further splitting the latter as sinh and cosh functions, one can get an expression for the periodic solution with an unknown parameter $c$.  This parameter can be derived by substituting the hyperbolic cot function form in the original evolutionary equation (for more details one may refer to \cite{Chow} and the references cited therein).  The above is indeed an imbricate solution.  In fact a theorem on imbricate series (see for example, Theorem 3.1 in ref. \cite{Book}) asserts that any periodic function $f(x)$ with period $L$ has two series representations.  If the usual Fourier series is 
\begin{eqnarray}
f(x)=\frac{\alpha}{2\pi}\sum_{n=-\infty}^{\infty}g(\alpha n)e^{i2\pi xn/L}, \ \alpha>0,
\end{eqnarray}
then the alternative expansion is the imbricate series
\begin{eqnarray}
f(x)=\sum_{m=-\infty}^{\infty}G\left(\frac{2\pi}{L\alpha}(x-mL)\right),
\end{eqnarray}
where $G(k)$ is the Fourier transform of $g(x)$, that is,
\begin{eqnarray}
G(k)=\frac{1}{2\pi}\int_{-\infty}^{\infty}g(x)e^{ikx}dx.
\end{eqnarray}
\par However, the imbricate series of rogue waves which we consider in the following will be different from the usual way of applying such a series which is the superposition of the solitary waves.  The important point here is that the breather solution (\ref{pct12}) is being constructed by the product of two imbricate series of rogue wave as shown below.  For mathematical simplicity we consider the constant $k=0$ hereafter (for $k\neq 0$, see below).  We consider Eq. (\ref{pct12}) in a more general form, that is
\begin{eqnarray}
q_1=&&\tau_1\exp(i(\sigma t+\phi))\left(1+b\sum_{n=-\infty}^{\infty}\frac{1}{i\alpha t+v(x)+n}\right)\nonumber\\
&&\times \left(1+b\sum_{n=-\infty}^{\infty}\frac{1}{i\alpha t-v(x)+n}\right),
\nonumber
\end{eqnarray}
\begin{eqnarray}
q_2=&&\tau_2\exp(i(\sigma t+\phi))\left(1+b\sum_{n=-\infty}^{\infty}\frac{1}{i\alpha t+v(x)+n}\right)\nonumber\\
&&\times \left(1+b\sum_{n=-\infty}^{\infty}\frac{1}{i\alpha t-v(x)+n}\right),
\label{pct13}
\end{eqnarray}
where $b$ is a constant, $\alpha$, $\sigma$ and $v(x)$ are all to be determined.  However, to derive the AB solution we consider only this series.  In the above expression we have grouped the spatial variable $x$ with the real part and the time variable $t$ with the imaginary part.  We have superposed the RW solutions in the $x$ direction. Using the trigonometric identity \cite{Integrals} $\cot\pi x=\frac{1}{\pi x}+\frac{x}{\pi}\sum_{n=-\infty}^\infty\frac{1}{n(x-n)}, \ n\neq 0$, we replace the infinite series by cot function and rewrite (\ref{pct13}) in a more compact form as
\begin{eqnarray}
q_1&=&\tau_1\exp(i(\sigma t+\phi))(1+b\pi \cot(\pi(v(x)+i\alpha t)))\nonumber\\&&\times(1-b\pi \cot(\pi(v(x)-i\alpha t))),\nonumber\\
q_2&=&\tau_2\exp(i(\sigma t+\phi))(1+b\pi \cot(\pi(v(x)+i\alpha t)))\nonumber\\&&\times(1-b\pi \cot(\pi(v(x)-i\alpha t))).
\label{pct14}
\end{eqnarray} 
\par Our task is to plug the expression (\ref{pct14}) in (\ref{cnls01}) and determine the parameters $\alpha$, $\sigma$ and the function $v(x)$ consistently.  To do so, in the first step, we replace the $\cot$ functions in (\ref{pct14}) as $\cos(\pi v(x)\pm i\pi\alpha t)/\sin(\pi v(x)\pm i\pi\alpha t)$ and substitute it in (\ref{cnls01}) and rewrite the equations in terms of $\sin(\pi v(x)\pm i\pi\alpha t)$ and $\cos(\pi v(x)\pm i\pi\alpha t)$ and their powers and products.  We then simplify these equations using suitable trigonometric identities and rearrange the resultant expressions in the variables $\cos(i\pi\alpha t)\sin(i\pi\alpha t)$ and their powers.  By doing so we have arranged the spatial variable to appear only in the coefficients in the resultant equation.  The final expressions for both the equations in (\ref{cnls01}) turn out to be one and the same.  As a result we proceed to determine the unknowns $\sigma$, $\mu$ and $v(x)$ from the single equation
\begin{eqnarray}
r_1\cos^6(i\pi\alpha t)+r_2\cos^5(i\pi\alpha t)&&\sin(i\pi\alpha t)+r_3\cos^4(i\pi\alpha t)\nonumber\\&&\hspace{-4.2cm}+r_4\cos^3(i\pi\alpha t)\sin(i\pi\alpha t)+r_5\cos^2(i\pi\alpha t)\nonumber\\&&\hspace{-4.2cm}+r_6\cos(i\pi\alpha t)\sin(i\pi\alpha t)+r_7=0,
\label{pct15}
\end{eqnarray}
where
\begin{eqnarray}
r_1&=&2\mu(\tau_1^2+\tau_2^2)(1+b^2\pi^2-b^4\pi^4-b^6\pi^6)-\sigma+b^2\pi^2\sigma,\nonumber\\
r_2&=&2b\pi(\sigma-2\mu(\tau_1^2+\tau_2^2)(1+b^2\pi^2)^2),\nonumber\\
r_3&=&2\mu(\tau_1^2+\tau_2^2)(-3\cos^2A-b^2\pi^2(1+\cos^2A)\nonumber\\
&&+b^4\pi^4(2-\cos^2A)+3b^6\pi^6\sin^2A)+3\sigma\cos^2A\nonumber\\
&&-b^2\pi^2\sigma(1+\cos^2A)-4b^2\pi^4v'^2(1-2\cos^2A)\nonumber\\
&&+2b\pi^2\alpha(1-2\cos^2A)+4b^2\pi^3v''\cos A\sin A,\nonumber\\
r_4&=&8\mu(\tau_1^2+\tau_2^2)b\pi(\cos^2A+b^2\pi^2(1+b^2\pi^2\sin^2A))\nonumber\\
&&+4b\pi(-\sigma\cos^2A+\pi^2v'^2(2\cos^2A-1)\nonumber\\
&&+\pi v''\cos A\sin A-\frac{1}{2}b\pi^2\alpha(1-2\cos^2A)),\nonumber\\
r_5&=&2\mu(\tau_1^2+\tau_2^2)(3\cos^4A-b^2\pi^2\cos^2A(2-3\cos^2A)\nonumber\\
&&+b^4\pi^4(-1+4\cos^2A-3\cos^4A)-3b^6\pi^6\sin^4A)\nonumber\\
&&-3\sigma\cos^4A+b^2\pi^2\sigma\cos^2A(2-\cos^2A)\nonumber\\
&&+2b^2\pi^4v'^2(1-8\cos^2A+4\cos^4A)+4b\pi^2\alpha\cos^4A\nonumber\\
&&-2b^2\pi^3v''\cos A\sin A(2\cos^2A+1),\nonumber\\
r_6&=&4\mu(\tau_1^2+\tau_2^2)b\pi(-\cos^4A+2b^2\pi^2\cos^2A\sin^2A\nonumber\\
&&-b^4\pi^4\sin^2A)+2b\pi\sigma\cos^4A+4b\pi^3v'^2\cos^2A\nonumber\\
&&\times(2\cos^2A-3)+4b\pi^2v''\cos^3A\sin A)\nonumber\\
&&+2b^2\pi^3\alpha\cos^2 A(1-2\cos^4A),\nonumber\\
r_7&=&2\mu(\tau_1^2+\tau_2^2)(-\cos^6A+3b^2\pi^2\cos^4A\sin^2A)\nonumber\\
&&-3b^4\pi^4\cos^2A\sin^4A+b^6\pi^6\sin^6A)+\sigma\cos^6A\nonumber\\
&&-b^2\pi^2\sigma\cos^4A\sin^2A+2b^2\pi^4v'^2\cos^2A(3-2\cos^2A)\nonumber\\
&&-2b\pi^2\alpha\cos^4A+2b^2\pi^3v''\cos^3A\sin A, \nonumber\\
A&=&\pi v(x).
\label{pct16}
\end{eqnarray}
 
\par To solve Eq. (\ref{pct15}) we equate the coefficients of various powers of $\cos(i\pi\alpha t) \sin(i\pi\alpha t)$ to zero.  This action yields a set of equations $r_i=0$, $i=1,2,..,7,$ involving the unknowns $\sigma$, $\alpha$ and $v(x)$.
We notice that the coefficient of sixth power of $\cos(i\pi\alpha t)$ gives $(r_1=0)$
\begin{eqnarray}
2\mu(\tau_1^2+\tau_2^2)(1+b^2\pi^2-b^4\pi^4-b^6\pi^6)-\sigma+b^2\pi^2\sigma=0,\nonumber\\
\label{pct17}
\end{eqnarray}
from which we can obtain the value of $\sigma$, that is
\begin{eqnarray}
\sigma=2\mu(\tau_1^2+\tau_2^2)(1+\pi^2b^2)^2.
\label{pct18}
\end{eqnarray}
\par The coefficient of $\cos^5(i\pi\alpha t)\sin(i\pi\alpha t)$ also provides the same expression for $\sigma$ as given in Eq.(\ref{pct18}).
Equating next the coefficients of $\cosh^3(i\pi\alpha t)$ and $\cos(i\pi\alpha t)\sin(i\pi\alpha t)$ to zero, we get $r_4=0$ and $r_6=0.$  Here we get two equations which contain the first and second derivatives of the unknown function $v(x)$, namely $v'$ and $v''$.  Solving these two equations algebraically, we find
\begin{eqnarray}
v'^2&=&b^2\mu(\tau_1^2+\tau_2^2)(1-b^2\pi^2\cot^2(2\pi v(x))),
\label{pct21}\\
v''&=&2\mu(\tau_1^2+\tau_2^2)\pi^3b^4\cot(2\pi v(x))\bigg(\frac{1+2\pi^2b^2}{\pi^2b^2}\nonumber\\&&-\frac{\alpha}{2\mu(\tau_1^2+\tau_2^2)\pi^2b^3}+\cot^2(2\pi v(x))\bigg).
\label{pct22}
\end{eqnarray}
From Eqs.(\ref{pct21}) and (\ref{pct22}) we determine $\alpha$ and $v(x)$ as follows.  Differentiating Eq.(\ref{pct21}) with respect to $x$ and then replacing the first and second derivatives of $v(x)$ which appear in this equation by (\ref{pct21}) and (\ref{pct22}) respectively and simplifying the resultant equation we find
\begin{eqnarray}
\alpha=2\mu(\tau_1^2+\tau_2^2)(1+\pi^2b^2)b.
\label{pct23}
\end{eqnarray}
\par To obtain $v(x)$ we integrate Eq.(\ref{pct21}) with respect to $x$.  This action leads us to
\begin{eqnarray}
v(x)=\frac{1}{2\pi}\arccos\left(\frac{1}{\sqrt{1+\pi^2b^2}}\cos(\sqrt{2\pi^2\alpha bx}+v_0)\right),\nonumber\\\label{pct24}
\end{eqnarray}
where $v_0$ is a constant of integration.  It is straight forward to check that $v(x)$ satisfies both the Eqs. (\ref{pct21}) and (\ref{pct22}) with $\alpha$ given by (\ref{pct23}).  On the other hand, considering the coefficients  of $\cos^2(i\pi\alpha t)$ and $\cos^4(i\pi\alpha t)$ and repeating the procedure outlined above we arrive at the same expressions for $\alpha$ and $v(x)$ which are given in (\ref{pct23}) and (\ref{pct24}) respectively.  
\par Finally, equating the coefficient of constant term to zero, $r_7=0$, we find that the resultant equation vanishes identically, with the expressions $v(x)$, $\alpha$ and $\sigma$ given above.  As a result we have obtained a compatible set of solutions for $\alpha$, $\sigma$ and $v(x)$ which satisfies all the equations given in (\ref{pct16}).  Now substituting the expressions of $\sigma$, $\alpha$ and $v(x)$ in the general form (\ref{pct14}) and after suitable rewriting, we obtain the AB solution in the form
\begin{eqnarray}
q_1&=&\tau_1(1+\pi^2b^2)\exp(i(2\mu(\tau_1^2+\tau_2^2)t+\phi))\nonumber\\&&\times(1-\frac{2\pi b}{1+\pi^2b^2}M),\nonumber\\
q_2&=&\tau_2(1+\pi^2b^2)\exp(i(2\mu(\tau_1^2+\tau_2^2)t+\phi))\nonumber\\&&\times(1-\frac{2\pi b}{1+\pi^2b^2}M),\label{pct25}\\
M&=&\frac{\pi b\cosh2\pi\alpha t+i\sinh2\pi\alpha t}{\cosh2\pi\alpha t-(1/\sqrt{1+\pi^2b^2})\cos(\sqrt{2\pi^2\alpha bx+v_0})}.\nonumber\\
\nonumber
\end{eqnarray}
\par We can note that this solution is periodic in the spatial direction and it grows exponentially fast in the initial stage from the time oscillatory background.  After reaching the maximum amplitude at a specific time, it decays exponentially again to the time oscillatory background.  These two stages can be called as growing and decaying mode solutions, respectively, as has been done by Tajiri and Watanabe for the case of the scalar NLS equation \cite{Tajiri}.  A typical AB solution for a suitable set of parametric values is shown in Fig. 2.
\par We also note here that the Akhmediev breather solution with $k \neq 0$ can also be constructed by the following imbricate series, namely
\begin{eqnarray}
q_1=&&\tau_1\exp(i(kx-(k^2+2\mu(\tau_1^2+\tau_2^2))t+\phi))\nonumber\\
&&\times \left(1+b\sum_{n=-\infty}^{\infty}\frac{1}{i\alpha t+v(z)+n}\right)\nonumber\\
&&\times \left(1+b\sum_{n=-\infty}^{\infty}\frac{1}{i\alpha t-v(z)+n}\right),
\nonumber
\end{eqnarray}
\begin{eqnarray}
q_2=&&\tau_2\exp(i(kx-(k^2+2\mu(\tau_1^2+\tau_2^2))t+\phi))\nonumber\\
&&\times \left(1+b\sum_{n=-\infty}^{\infty}\frac{1}{i\alpha t+v(z)+n}\right)\nonumber\\
&&\times \left(1+b\sum_{n=-\infty}^{\infty}\frac{1}{i\alpha t-v(z)+n}\right),
\label{pc13}
\end{eqnarray}
where $z=x-2kt$.  Substituting this expression into (\ref{cnls01}) and repeating the procedure outlined above one can obtain the Akhmediev breather with $k \neq 0$. 
%%%%%%%%%%%%%%%%%%%%%%%%%%%%%%%%%%%%%%%%%%%%%%%%%%%%%%%%%%%%%%%%%%%%%%%%%%%%%%%%%%%%%%%%%%%%%%%%%%%%%%%%%%%%%%%%%%%%%%%%%%%%%%%%%%%%%%%

\section{MS from RW}
\label{RWtoMS}
Next we construct the MS solution from the RW solution.  To do so we again rewrite the RW solution in (\ref{pct8}) in a slightly different factorized form
\begin{eqnarray}
q_1=&&\tau_1\exp(i(kx-(k^2-2\mu(\tau_1^2+\tau_2^2))t))\nonumber\\
&&\hspace{-.8cm}\times\bigg(1+\frac{i}{-2\mu(\tau_1^2+\tau_2^2)t+i\frac{1}{2}\sqrt{1+4\mu(\tau_1^2+\tau_2^2)(x-2kt)^2}}\bigg)\nonumber\\
&&\hspace{-.8cm}\times\bigg(1+\frac{i}{-2\mu(\tau_1^2+\tau_2^2)t-i\frac{1}{2}\sqrt{1+4\mu(\tau_1^2+\tau_2^2)(x-2kt)^2}}\bigg),\nonumber
\end{eqnarray}
\begin{eqnarray}
q_2=&&\tau_2\exp(i(kx-(k^2-2\mu(\tau_1^2+\tau_2^2))t))\nonumber\\
&&\hspace{-.8cm}\times\bigg(1+\frac{i}{-2\mu(\tau_1^2+\tau_2^2)t+i\frac{1}{2}\sqrt{1+4\mu(\tau_1^2+\tau_2^2)(x-2kt)^2}}\bigg)\nonumber\\
&&\hspace{-.8cm}\times\bigg(1+\frac{i}{-2\mu(\tau_1^2+\tau_2^2)t-i\frac{1}{2}\sqrt{1+4\mu(\tau_1^2+\tau_2^2)(x-2kt)^2}}\bigg).\nonumber\\
\label{pc12}
\end{eqnarray} 
We can write this equation in the following general form with $k=0$,
\begin{eqnarray}
q_1=&&\tau_1\exp(i(\zeta t+\phi))\left(1+ih\sum_{n=-\infty}^{\infty}\frac{1}{\kappa t+i\varrho(x)+n}\right)\nonumber\\
&&\times \left(1+ih\sum_{n=-\infty}^{\infty}\frac{1}{\kappa t-i\varrho(x)+n}\right),
\nonumber
\end{eqnarray}
\begin{eqnarray}
q_2=&&\tau_2\exp(i(\zeta t+\phi))\left(1+ih\sum_{n=-\infty}^{\infty}\frac{1}{\kappa t+i\varrho(x)+n}\right)\nonumber\\
&&\times \left(1+ih\sum_{n=-\infty}^{\infty}\frac{1}{\kappa t-i\varrho(x)+n}\right),
\label{pct26}
\end{eqnarray}
where the function $\varrho(x)$ and the parameters $\kappa$ and $\zeta$ are to be determined.  Here we have superposed the RW in the temporal direction.  We have also grouped the temporal variable with the real part and the spatial variable with the imaginary part.  We identify the infinite series with the $\cot$ hyperbolic function \cite{Integrals}, $\coth\pi x=\frac{1}{\pi x}-\frac{ix}{\pi}\sum_{n=-\infty}^\infty\frac{1}{n(x-in)}, n\neq 0$, and rewrite the above expression as 
\begin{eqnarray}
q_1&=&\tau_1\exp(i(\zeta t+\phi))(1+h\pi \coth(\pi(\varrho(x)-i\kappa t)))\nonumber\\&&\times(1-h\pi \coth(\pi(\varrho(x)+i\kappa t))),\nonumber\\
q_2&=&\tau_2\exp(i(\zeta t+\phi))(1+h\pi \coth(\pi(\varrho(x)-i\kappa t)))\nonumber\\&&\times(1-h\pi \coth(\pi(\varrho(x)+i\kappa t))).
\label{pct27}
\end{eqnarray} 
\par As we did previously, we split the $\cot$ hyperbolic function as $\cosh(\pi\varrho(x)\pm i\pi\kappa t)/\sinh(\pi\varrho(x)\pm i\pi\kappa t)$. We then substitute the expression (\ref{pct27}) into the CNLS equations (\ref{cnls01}) and rewrite the latter in terms of $\cosh(\pi\varrho(x)\pm i\pi\kappa t)/\sinh(\pi\varrho(x)\pm i\pi\kappa t)$.  As before, we simplify this equation further by imposing trigonometric identities and arrive at an equation which is in powers of $\cosh(i\pi\kappa t)$ $\sinh(i\pi\kappa t)$ and their products.  In this case also we find that both the equations in (\ref{cnls01}) provide the same expression.  As a result we consider only the following equation to determine the unknown parameters, that is
\begin{eqnarray}
z_1\cosh^6(i\pi\kappa t)+z_2\cosh^5(i\pi\kappa t)&&\sinh(i\pi\kappa t)+z_3\cosh^4(i\pi\kappa t)\nonumber\\
&&\hspace{-4.2cm}+z_4\cosh^3(i\pi\kappa t)\sinh(i\pi\kappa t)+z_5\cosh^2(i\pi\kappa t)\nonumber\\&&\hspace{-4.2cm}+z_6\cosh(i\pi\kappa t)\sinh(i\pi\kappa t)+z_7=0,
\label{pct28}
\end{eqnarray}
where
\begin{eqnarray}
z_1&=&2\mu(\tau_1^2+\tau_2^2)(-1+h^2\pi^2-h^4\pi^4-h^6\pi^6)+\zeta+h^2\pi^2\zeta,\nonumber\\
z_2&=&4h\pi\mu(\tau_1^2+\tau_2^2)(1-h^2\pi^2)^2-2h\pi\zeta,\nonumber\\
z_3&=&2\mu(\tau_1^2+\tau_2^2)(3\cosh^2A-h^2\pi^2(1+\cosh^2A)\nonumber\\
&&+h^4\pi^4(\cosh^2A-2)-3h^6\pi^6\sinh^2A)-3\zeta\cosh^2A\nonumber\\
&&-h^2\pi^2\zeta(1+\cosh^2A)+4h^2\pi^4\varrho'^2(1-2\cosh^2A)\nonumber\\
&&+2h\pi^2\kappa(2\cosh^2A-1)-4h^2\pi^3\varrho''\cosh A\sinh A,\nonumber\\
z_4&=&8\mu(\tau_1^2+\tau_2^2)h\pi(-\cosh^2A+h^2\pi^2+h^4\pi^4\sinh^2A)\nonumber\\
&&+4h\pi\zeta\cosh^2A+2h\pi^3\varrho'^2(-1+2\cosh^2A)\nonumber\\
&&+4h\pi^2\varrho''\cosh A\sinh A+2h^2\pi^3\kappa(1-2\cosh^2A),\nonumber\\
z_5&=&2\mu(\tau_1^2+\tau_2^2)(-3\cosh^4A-h^2\pi^2\cosh^2A(2-3\cosh^2A)\nonumber\\
&&+h^4\pi^4(1-4\cosh^2A+3\cosh^4A-3h^6\pi^6\cosh^2A\nonumber\\
&&\times\sinh^2A))+3\zeta\cosh^4A-4h\pi^2\kappa\cosh^4A\nonumber\\
&&+2h^2\pi^2\zeta\cosh^2A(1-2\cosh^2A)-2h^2\pi^4\varrho'^2\nonumber\\
&&\times(1-8\cosh^2A+4\cosh^4A)-2h^2\pi^3\varrho''\cosh A\nonumber\\
&&\times\sinh A(2\cosh^2A+1),\nonumber\\
z_6&=&4\mu(\tau_1^2+\tau_2^2)h\pi(\cosh^4A-2h^2\pi^2\cosh^2A\sinh^2A\nonumber\\
&&+h^4\pi^4\sinh^4A)-2h\pi\zeta\cosh^4A+4h\pi^3\varrho'^2\cosh^2A\nonumber\\
&&\times(2\cosh^2A-3)-4h\pi^2\varrho''\cosh^3A\sinh A\nonumber\\
&&-2h^2\pi^3\kappa\cosh^2 A(1-2\cosh^2A),\nonumber\\
z_7&=&2\mu(\tau_1^2+\tau_2^2)(\cosh^6A-3h^2\pi^2\cosh^4A\sinh^2A\nonumber\\
&&+3h^4\pi^4\cosh^2A\sinh^4A-h^6\pi^6\sinh^6A)-\zeta\cosh^6A\nonumber\\
&&+h^2\pi^2\zeta\cosh^4A\sinh^2A+2h^2\pi^4\varrho'^2\cosh^2A\nonumber\\
&&\times(2\sinh^2A-1)+2h\pi^2\kappa\cosh^4A\nonumber\\
&&-2h^2\pi^3\varrho''\cosh^3A\sinh A,\nonumber\\
A&=&\pi\varrho(x)
\label{pct29}
\end{eqnarray}
\par Equating the various powers of $\cosh(i\pi\kappa t)\sinh(i\pi\kappa t)$ to zero, we obtain $z_i=0$, $i=1,2..,7$.  We then solve these equations and determine $\zeta$, $\varrho(x)$ and $\kappa$ as follows.  The coefficient of $\cosh^6(i\pi\kappa t)$ gives
\begin{eqnarray}
2\mu(\tau_1^2+\tau_2^2)(-1+h^2\pi^2+h^4\pi^4-h^6\pi^6)+\zeta+\zeta h^2\pi^2=0,\nonumber\\
\label{pct30}
\end{eqnarray}
from which we fix
\begin{eqnarray}
\zeta=2\mu(\tau_1^2+\tau_2^2)(1-\pi^2h^2)^2.
\label{pct29}
\end{eqnarray}
We also obtain the same expression for $\zeta$ by equating the coefficient of $\cosh^5(i\pi\kappa t)\sinh(i\pi\kappa t)$ to zero.  We proceed to consider the coefficients of  $\cosh^3(i\pi\kappa t)\sinh(i\pi\kappa t)$ and  $\cosh(i\pi\kappa t)\sinh(i\pi\kappa t)$, namely $z_4=0$ and $z_6=0.$  We consider these two expressions to determine the unknown $\varrho(x)$. 
Solving these two equations algebraically, we find
\begin{eqnarray}
\varrho'^2&=&h^2\mu(\tau_1^2+\tau_2^2)(1-h^2\pi^2\coth^2(2\pi\varrho(x))),\nonumber\\
\label{pct32}
\varrho''&=&2\mu(\tau_1^2+\tau_2^2)\pi^3h^4\coth(2\pi\varrho(x))\bigg(\frac{1-2\pi^2h^2}{\pi^2h^2}\nonumber\\&&+\frac{\kappa}{2\mu(\tau_1^2+\tau_2^2)\pi^2h^3}+\coth^2(2\pi\varrho(x))\bigg).
\label{pct33}
\end{eqnarray}
We solve these two equations in the same manner as we did previously.  Our result shows that
\begin{eqnarray}
\kappa=-2\mu(\tau_1^2+\tau_2^2)(1-\pi^2h^2)h
\label{pct34}
\end{eqnarray} 
and
\begin{eqnarray}
\varrho(x)=\frac{1}{2\pi}\cosh^{-1}\bigg(\frac{1}{\sqrt{1-\pi^2h^2}}\cosh(\sqrt{-2\pi^2\kappa hx}+\varrho_0)\bigg),\nonumber\\
\label{pct35}
\end{eqnarray}
where $\rho_0$ is a constant.  We can also obtain the same expression for $\varrho(x)$ and $\kappa$ from the coefficients of $\cosh^2(i\pi\kappa t)$ and $\cosh^4(i\pi\kappa t)$ by solving the resultant equations algebraically in the same manner.  
\par Inserting the obtained expressions of $\zeta$, $\varrho$ and $\kappa$ in the final determining equation we find that it vanishes trivially.  With these expressions, the general form of (\ref{pct26}) now becomes
\begin{eqnarray}
q_1&=&\tau_1(1-\pi^2h^2)\exp(i(2\mu(\tau_1^2+\tau_2^2)t+\phi))\nonumber\\&&\times(1+\frac{2\pi h}{1-\pi^2h^2}M),\nonumber\\
q_2&=&\tau_2(1-\pi^2h^2)\exp(i(2\mu(\tau_1^2+\tau_2^2)t+\phi))\nonumber\\&&\times(1+\frac{2\pi h}{1-\pi^2h^2}M),\label{pct36}\\
M&=&\frac{\pi h\cos2\pi\kappa t-i\sin2\pi\kappa t}{\cos2\pi\kappa t-(1/\sqrt{1-\pi^2h^2})\cosh(\sqrt{-2\pi^2\kappa hx+c})},\nonumber\\
\nonumber
\end{eqnarray}
which is nothing but the Ma breather solution.  This solution is periodic in the temporal direction and localized in space.  It grows and decays recurrently in time oscillate background as in the case of NLS equation \cite{Tajiri}.  The Ma breather solution of CNLS equations for a set of parametric values is shown in Fig. 3.
\par The RW solution can also be obtained as the limiting case of Ma breathers.  This can be done by imposing the limit $h\to 0$ and incorporating the Taylor series expansion.
\par We note here that the Ma breather solution with $k \neq 0$ can also be constructed by the following imbricate series,
\begin{eqnarray}
q_1=&&\tau_1\exp(i(kx-(k^2-2\mu(\tau_1^2+\tau_2^2))t))\nonumber\\
&&\times \left(1+ih\sum_{n=-\infty}^{\infty}\frac{1}{\kappa t+i\varrho(z)+n}\right)\nonumber\\
&&\times \left(1+ih\sum_{n=-\infty}^{\infty}\frac{1}{\kappa t-i\varrho(z)+n}\right),
\nonumber
\end{eqnarray}
\begin{eqnarray}
q_2=&&\tau_2\exp(i(kx-(k^2+2\mu(\tau_1^2+\tau_2^2))t+\phi))\nonumber\\
&&\times \left(1+ih\sum_{n=-\infty}^{\infty}\frac{1}{\kappa t+i\varrho(z)+n}\right)\nonumber\\
&&\times \left(1+ih\sum_{n=-\infty}^{\infty}\frac{1}{\kappa t-i\varrho(z)+n}\right),
\label{p13}
\end{eqnarray}
where $z=x-2kt$.  Substituting this expression into (\ref{cnls01}), and repeating the procedure given above, one can obtain the Ma soliton with $k \neq 0$. 
%%%%%%%%%%%%%%%%%%%%%%%%%%%%%%%%%%%%%%%%%%%%%%%%%%%%%%%%%%%%%%%%%%%%%%%%%%%%%%%%%%%%%%%%%%%%%%%%%%%%%%%%%%%%%%%%%%%%%%%%%%%%%%%%
\section{GB{\sf s} as imbricate series of RW{\sf s}} 
\label{RWtoGB}
It is very difficult to derive the GB solution from the RW solution in the same fashion as we did in the previous two cases.  This is mainly because in the present analysis we have to include two arbitrary functions (both of which are functions of $t$ and $x$), one with real part and another with imaginary part in the imbricate series.  The determining equations which come out from the imbricate series are difficult to solve unlike the earlier two cases.  To overcome this difficulty we adopt the following methodology.  We show that the absolute square of the modulus of RW solution of (\ref{cnls01}) can be written in terms of the second derivative of a logarithmic function which involves product of two imbricate series (see Eq.(\ref{pct38}) given below).  We then rewrite this expression in a more compact form which involves trigonometric functions which also contain these two arbitrary functions.  Unlike the earlier two cases we do not substitute this series into (\ref{cnls01}) and determine these two unknown arbitrary functions (as it is very difficult to solve the determining equation).  Instead, we also rewrite the absolute square of the modulus of the general breather solution as the second derivative of a logarithmic function which contains product of two functions (see Eq. (\ref{pct42}) given below).  At this stage since both the GB expression and the RW solution have been written in the same form.  We compare the arguments inside the logarithmic function and fix the exact expression of the two unknown arbitrary functions.  This in turn confirms that the GB can also be derived from the RWs.  In the following, we present the exact mathematical details of this procedure.  
\par To begin with we rewrite the RW solution (\ref{pct11}) in the form  
\begin{eqnarray}
|q_1|^2=\tau_1^2(1-Q)(1-Q^*), \quad |q_2|^2=\tau_2^2(1-Q)(1-Q^*),\nonumber\\\label{pc36}
\end{eqnarray}
where $Q$ is the same expression given in Eq. (\ref{pct11}) and the star denotes complex conjugate of it.  The above expressions can be rewritten as the second derivative of a logarithimic function, namely
\begin{eqnarray}
|q_1|^2&=&\tau_1^2-\frac{\tau_1^2}{\mu(\tau_1^2+\tau_2^2)}\frac{\partial^2}{\partial x^2}\ln\bigg(\frac{1}{S} \times \frac{1}{T}\bigg),\nonumber\\
|q_2|^2&=&\tau_2^2-\frac{\tau_2^2}{\mu(\tau_1^2+\tau_2^2)}\frac{\partial^2}{\partial x^2}\ln\bigg(\frac{1}{S} \times \frac{1}{T}\bigg),\label{pct37}
\end{eqnarray} 
where
\begin{eqnarray}
S=\bigg(\frac{1}{2}\sqrt{1+4\mu(\tau_1^2+\tau_2^2)(x-2kt)^2}+2i\mu(\tau_1^2+\tau_2^2)t\bigg)^2,\nonumber\\
T=\bigg(\frac{1}{2}\sqrt{1+4\mu(\tau_1^2+\tau_2^2)(x-2kt)^2}-2i\mu(\tau_1^2+\tau_2^2)t\bigg)^2.\nonumber
\end{eqnarray}
\par Now we consider Eq. (\ref{pct37}) in a more general form as
\begin{eqnarray}
|q_1|^2&=&\tau_1^2-\frac{\tau_1^2}{\mu(\tau_1^2+\tau_2^2)}\nonumber\\
&&\times\frac{\partial^2}{\partial x^2}\ln\bigg(\sum_{n=-\infty}^\infty\frac{1}{(\phi(x,t)-i\psi(x,t)-n)^2}\nonumber\\&&\times\sum_{n=-\infty}^\infty\frac{1}{(\phi(x,t)+i\psi(x,t)-n)^2}\bigg),\nonumber\\
|q_2|^2&=&\tau_2^2-\frac{\tau_2^2}{\mu(\tau_1^2+\tau_2^2)}\nonumber\\
&&\times\frac{\partial^2}{\partial x^2}\ln\bigg(\sum_{n=-\infty}^\infty\frac{1}{(\phi(x,t)-i\psi(x,t)-n)^2}\nonumber\\&&\times\sum_{n=-\infty}^\infty\frac{1}{(\phi(x,t)+i\psi(x,t)-n)^2}\bigg),\label{pct38}
\end{eqnarray}
where $\phi(x,t)$ and $\psi(x,t)$ are arbitrary functions of $x$ and $t$ which need to be determined.  In the above expression we considered the superposition of RWs in both space and time directions.  Using the trigonometric identity \cite{Integrals} $\csc^2(\pi x)=\frac{1}{\pi^2}\sum_{k=-\infty}^\infty\frac{1}{x-k^2}$, the above expression can be rewritten in the form, 
\begin{eqnarray}
|q_1|^2&=&\tau_1^2-\frac{\tau_1^2}{\mu(\tau_1^2+\tau_2^2)}\frac{\partial^2}{\partial x^2}\ln[\pi^2\csc^2(\pi(\phi-i\psi))\nonumber\\&&\times\pi^2\csc^2(\pi(\phi+i\psi))],\nonumber\\
|q_2|^2&=&\tau_2^2-\frac{\tau_2^2}{\mu(\tau_1^2+\tau_2^2)}\frac{\partial^2}{\partial x^2}\ln[\pi^2\csc^2(\pi(\phi-i\psi))\nonumber\\&&\times\pi^2\csc^2(\pi(\phi+i\psi))].\label{pct39}
\end{eqnarray}
We further simplify the expression on the right hand side by using the relation $\pi^4\csc^2(\pi(\phi-i\psi))\csc^2(\pi(\phi+i\psi))$ = $\frac{4\pi^4}{\cosh 2\pi\psi-\cos 2\pi\phi}$.  As a result Eq. (\ref{pct39}) can be brought to the form  
\begin{eqnarray}
|q_1|^2&=&\tau_1^2+\frac{\tau_1^2}{\mu(\tau_1^2+\tau_2^2)}\frac{\partial^2}{\partial x^2}\ln[\cosh2\pi\psi-\cos2\pi\phi],\nonumber\\
|q_2|^2&=&\tau_2^2+\frac{\tau_2^2}{\mu(\tau_1^2+\tau_2^2)}\frac{\partial^2}{\partial x^2}\ln[\cosh2\pi\psi-\cos2\pi\phi].\nonumber\\\label{pct40}
\end{eqnarray}
\par As we mentioned in the beginning of this section we do not substitute this solution in (\ref{cnls01}) and determine the form of $\psi$ and $\phi$.  Instead of this we compare the expression (\ref{pct40}) with the GB which is rewritten in the same form.  For this purpose, we rewrite the GB solution (\ref{pc4}) with (\ref{pct5}) in the form,
\begin{eqnarray}
|q_1|^2=\tau_1^2+\frac{\tau_1^2}{\mu(\tau_1^2+\tau_2^2)}\frac{\partial^2}{\partial x^2}\ln f,\nonumber\\
|q_2|^2=\tau_2^2+\frac{\tau_1^2}{\mu(\tau_1^2+\tau_2^2)}\frac{\partial^2}{\partial x^2}\ln f, \label{pct41}
\end{eqnarray}
where $f=1+2e^{\eta_R}\cos\eta_I+ae^{2\eta_R}$, $\eta_R=p_Rx-\Omega_Rt+\eta_R^0$ and $\eta_I=p_I-\Omega_It+\eta_I^0$.  To compare this with the one derived from the RW solution we rewrite $f$ as $f=2e^{\eta_R}(\sqrt{a}\cosh(\eta_R+\sigma)-\cos(\eta_I+\theta))$ with the $\eta_R$ and $\eta_I$ as given above.  The resultant expression now turns out to be
\begin{eqnarray}
|q_1|^2&=&\tau_1^2+\frac{\tau_1^2}{\mu(\tau_1^2+\tau_2^2)}\frac{\partial^2}{\partial x^2}\ln[\sqrt{a}\cosh(p_Rx-\Omega_Rt+\sigma)\nonumber\\&&-\cos(p_Ix-\Omega_It+\theta)],\nonumber\\
|q_2|^2&=&\tau_2^2+\frac{\tau_2^2}{\mu(\tau_1^2+\tau_2^2)}\frac{\partial^2}{\partial x^2}\ln[\sqrt{a}\cosh(p_Rx-\Omega_Rt+\sigma)\nonumber\\&&-\cos(p_Ix-\Omega_It+\theta)],\label{pct42}
\end{eqnarray}
where $\sigma=\eta_R^0+\frac{1}{2}\ln a$ and $\theta=\eta_I^0+\pi$.  
\par Now let us compare the two expressions $|q_1|^2$ and $|q_2|^2$, the one derived from RW solutions (vide Eq. (\ref{pct40})) and the other derived from the GB solution (vide Eq. (\ref{pct42})).  Doing so we find
\begin{eqnarray}
\cosh 2\pi\psi&=&\sqrt{a}\cosh(p_Rx-\Omega_Rt+\sigma),\nonumber\\
\cos 2\pi\phi&=&\cos(p_Ix-\Omega_It+\theta),
\end{eqnarray}
or
\begin{eqnarray}
\cosh 2\pi\psi&=&\cosh(p_Rx-\Omega_Rt+\sigma),\nonumber\\
\cos 2\pi\phi&=&\frac{1}{\sqrt{a}}\cos(p_Ix-\Omega_It+\theta).
\end{eqnarray}
From these two sets of equations we find two different expressions for $\psi$ and $\phi$, namely
\begin{eqnarray}
\psi&=&\frac{1}{2\pi}\ln(\sqrt{a}\cosh(p_Rx-\Omega_Rt+\sigma)\nonumber\\&&+\sqrt{a\cosh^2(p_Rx-\Omega_Rt+\sigma)-1}),\nonumber\\
\phi&=&\frac{1}{2\pi}(p_Ix-\Omega_It+\theta)\label{pct43}
\end{eqnarray}
and
\begin{eqnarray}
\psi&=&-\frac{1}{2\pi}(p_Rx-\Omega_Rt+\sigma),\nonumber\\
\phi&=&\frac{1}{2\pi}\arccos\left(\frac{1}{\sqrt{a}}\cos(p_Ix-\Omega_It+\theta)\right).\label{pct44}
\end{eqnarray}
An exact imbricate series of RW solution for breather solutions of CNLS equations can be displayed by substituting (\ref{pct43}) or (\ref{pct44}) into (\ref{pct38}).  The solution is periodic in both space and time. 

%%%%%%%%%%%%%%%%%%%%%%%%%%%%%%%%%%%%%%%%%%%%%%%%%%%%%%%%%%%%%%%%%%%%%%%%%%%%%%%%%%%%%%%%%%%%%%%%%%%%%%%%%%%%%%%%%%%%%%%%%%%%%%%%%%%%%
\section{Conclusion}
\label{conclusions}
During the past five years or so rogue wave solutions have been studied intensively in different physical contexts and several applications have been proposed.  Rogue waves in an array of optical wave-guides is one of the examples \cite{conclusion1}.  Optical rogue wave has already been suggested for application in enhancing supercontinuum generation and several authors have elucidated key aspects of the underlying nonlinear dynamical processes \cite{conclusion2}.  In this work, we have constructed a class of nonlinear waves, namely GB, AB, MS and RW for the well known two coupled NLSEs (\ref{cnls01}).  To derive these solutions we followed two different paths.  By following the conventional procedure we first brought out the explicit form of a GB solution from which we derived the other forms of rational solutions, namely AB, MS and RW solutions.  We then deviated from this conventional approach and derived AB, MS and GB from  the RW solution as the starting point.  The expressions obtained in both the directions match with each other.  Our study on the coupled NLSEs will be useful in the study of rogue waves in birefringent optical fibers, multi-component Bose-Einstein condensates, multi-component plasmas and so on.  We also hope to derive higher order breather solutions in both the directions discussed in this paper by extending the procedure.

\section*{Acknowledgements}
NVP wishes to thank the University Grants Commission (UGC-RFSMS), 
Government of India, for providing a Research Fellowship. The work of MS forms part of a research project sponsored by NBHM, Government of India and while the work of ML forms part of an IRHPA project and a Ramanna Fellowship project 
of ML, sponsored by the Department of Science and Technology (DST), Government of India. ML 
also acknowledges the financial support under a DAE Raja Ramanna Fellowship.

\end{document}